\documentclass[a4paper,11pt]{article}

\usepackage{jinstpub} 

\usepackage[modulo]{lineno}

\usepackage{xparse,mathtools}

\DeclarePairedDelimiterX{\rvect}[1]{[}{]}{\,\makervect{#1}\,}

\ExplSyntaxOn
\NewDocumentCommand{\makervect}{m}
 {
  \seq_set_split:Nnn \l_tmpa_seq { , } { #1 }
  \begin{matrix}
  \seq_use:Nn \l_tmpa_seq { & }
  \end{matrix}
 }
\ExplSyntaxOff

\usepackage[utf8]{inputenc}
\usepackage{algorithm2e}
\usepackage{graphicx}
\usepackage{subfig}
\usepackage[percent]{overpic}
\usepackage{pict2e}
\usepackage{setspace,lipsum}
\usepackage{epsfig}
\usepackage{color}
\usepackage{amsmath,amsfonts} 
\usepackage[version=3]{mhchem}
\usepackage{hyperref}
\usepackage{enumitem}
\usepackage{siunitx}
\usepackage{xfrac}
\usepackage[useregional]{datetime2}
\usepackage[dvipsnames]{xcolor}
\usepackage{wrapfig}
\usepackage{float}
\usepackage{booktabs}
\usepackage{todonotes}
\usepackage[nolist]{acronym}
\usepackage{xspace}
\usepackage[capitalize]{cleveref}

\DeclareSIUnit\kminv{\km\tothe{-1}}
\DeclareSIUnit\day{day}
\DeclareSIUnit\dayinv{\day\tothe{-1}}

\newcommand{\gnn}{\textsc{dynedge}\xspace}
\newcommand{\retro}{\textsc{retro}\xspace}

\newcommand{\trackvscascade}{$\mathcal{T}/\mathcal{C}$\xspace}
\newcommand{\nuvsmu}{$\nu / \mu$\xspace}
\newcommand{\tc}{\ensuremath{\mathcal{T}/\mathcal{C}}\xspace}

\makeatletter
\newcommand*{\shifttext}[2]{%
  \settowidth{\@tempdima}{#2}%
  \makebox[\@tempdima]{\hspace*{#1}#2}%
}
\makeatother

\definecolor{dark-grey}{gray}{0.4}
\definecolor{darkgreen}{rgb}{0,0.5,0}
\definecolor{orange}{rgb}{0.93, 0.53, 0.18}
\begin{document}

\begin{acronym}[TDMA]
\acro{BDT}{Boosted Decision Tree}
\acro{SK}{Super-Kamiokande}
\acro{HK}{Hyper-Kamiokande}
\acro{SM}{Standard Model}
\acro{BSM}{Beyond the Standard Model}
\acro{MC}{Monte Carlo}
\acro{ML}{Machine Learning}
\acro{MLP}{Multilayer Perceptron}
\acro{NMO}{Neutrino Mass Ordering}
\acro{CNN}{Convolutional Neural Network}
\acro{GNN}{Graph Neural Network}
\acro{GAN}{Generative Adversarial Network}
\acro{GPU}{Graphics Processing Unit}
\acro{CPU}{Central Processing Unit}
\acro{CC}{charged current}
\acro{NC}{neutral current}
\acro{DOM}{Digital Optical Module}
\acro{MTOM}{Muon Tomography Optical Module}
\acro{SiPM}{Silicon Photomultiplier}
\acro{PMT}{Photomultiplier Tube}
\acro{MCP}{Microchannel Plate}
\acro{DAQ}{Data Acquisition}
\acro{CR}{Cosmic Ray}
\acro{PMNS}{Pontecorvo-Maki-Nakagawa-Sakata}
\acro{MSW}{Mikheyev-Smirnov-Wolfenstein}
\acro{DUNE}{Deep Underground Neutrino Experiment}
\acro{JUNO}{Jiangmen Underground Neutrino Observatory}
\acro{FPGA}{Field-Programmable Gate Array}
\acro{IC}{IceCube}
\acro{DC}{DeepCore}
\acro{NN}{Neural Network}
\acro{ROC}{Receiver Operating Characteristics}
\acro{QE}{quantum efficiency}
\acro{BCE}{Binary Cross Entropy}
\acro{AUC}{Area Under the Curve}
\acro{knn}[\emph{k}-nn]{\emph{k}-nearest neighbors}
\end{acronym}

\title{Graph Neural Networks for Low-Energy Event Classification \& Reconstruction in IceCube}

\author[16]{R. Abbasi,}
\author[62]{M. Ackermann,}
\author[17]{J. Adams,}
\author[24]{N. Aggarwal,}
\author[11]{J. A. Aguilar,}
\author[21]{M. Ahlers,}
\author[52]{M. Ahrens,}
\author[22]{J.M. Alameddine,}
\author[30]{A. A. Alves Jr.,}
\author[42]{N. M. Amin,}
% ...
\affiliation[11]{Universit{\'e} Libre de Bruxelles, Science Faculty CP230, B-1050 Brussels, Belgium}
% ...
\affiliation[16]{Department of Physics, Loyola University Chicago, Chicago, IL 60660, USA}
\affiliation[17]{Dept. of Physics and Astronomy, University of Canterbury, Private Bag 4800, Christchurch, New Zealand}
% ...
\affiliation[21]{Niels Bohr Institute, University of Copenhagen, DK-2100 Copenhagen, Denmark}
\affiliation[22]{Dept. of Physics, TU Dortmund University, D-44221 Dortmund, Germany}
% ...
\affiliation[24]{Dept. of Physics, University of Alberta, Edmonton, Alberta, Canada T6G 2E1}
% ...
\affiliation[30]{Karlsruhe Institute of Technology, Institute for Astroparticle Physics, D-76021 Karlsruhe, Germany }
% ...
\affiliation[42]{Bartol Research Institute and Dept. of Physics and Astronomy, University of Delaware, Newark, DE 19716, USA}
% ...
\affiliation[52]{Oskar Klein Centre and Dept. of Physics, Stockholm University, SE-10691 Stockholm, Sweden}
% ...
\affiliation[62]{DESY, D-15738 Zeuthen, Germany}
% ...

\emailAdd{analysis@icecube.wisc.edu}

\setlength\linenumbersep{15pt}
\renewcommand\linenumberfont{\normalfont\tiny\sffamily\color{black}}

\abstract{
IceCube, a cubic-kilometer array of optical sensors built to detect atmospheric and astrophysical neutrinos between 1 GeV and 1 PeV, is deployed 1.45 km to 2.45 km below the surface of the ice sheet at the South Pole. The classification and reconstruction of events from the in-ice detectors play a central role in the analysis of data from IceCube.  Reconstructing and classifying events is a challenge due to the irregular detector geometry, inhomogeneous scattering and absorption of light in the ice and, below 100 GeV, the relatively low number of signal photons produced per event.
To address this challenge, it is possible to represent IceCube events as point cloud graphs and use a Graph Neural Network (GNN) as the classification and reconstruction method. The GNN is capable of distinguishing neutrino events from cosmic-ray backgrounds, classifying different neutrino event types, and reconstructing the deposited energy, direction and interaction vertex. Based on simulation, we provide a comparison in the 1\,GeV--100\,GeV energy range to the current state-of-the-art maximum likelihood techniques used in current IceCube analyses, including the effects of known systematic uncertainties.
For neutrino event classification, the GNN increases the signal efficiency by 18\% at a fixed background rate, compared to current IceCube methods. Alternatively, the GNN offers a reduction of the background (i.e.\ false positive) rate by over a factor 8 (to below half a percent) at a fixed signal efficiency.
For the reconstruction of energy, direction, and interaction vertex, the resolution improves by an average of 13\%--20\% compared to current maximum likelihood techniques in the energy range of 1\,GeV--30\,GeV.
The GNN, when run on a GPU, is capable of processing IceCube events at a rate nearly double of the median IceCube trigger rate of 2.7~kHz, which opens the possibility of using low energy neutrinos in online searches for transient events.
}




\date{\today}
\maketitle


\newpage
\section{Introduction}

\subsection{The IceCube Detector}

\begin{figure}[h!]
    \begin{center}
    \includegraphics[width=0.6\textwidth]{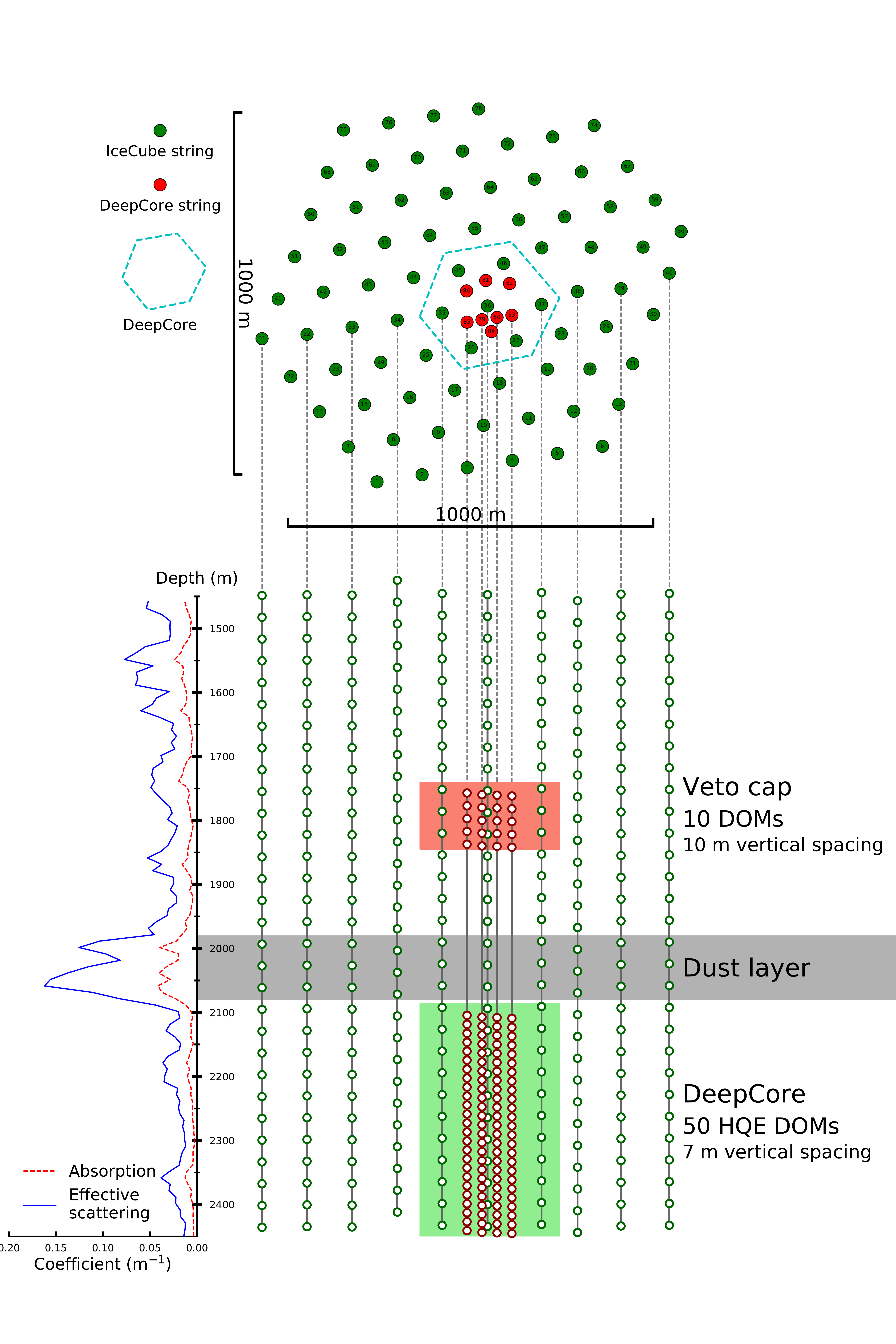}
    \caption{Top and side view of IceCube detector \cite{DeepCore}. Ice properties as a function of depth is shown on the left. The in-ice dust layer is marked in gray. The dust layer is a layer in the ice with dust impurities and therefore reduced optical qualities. DeepCore is placed under the dust layer and shown in green. High Quantum Efficiency (HQE) DOMs are placed under the dust layer in the DeepCore region.}
    \label{fig:deepcore_illustration}
    \end{center}
\end{figure}

The IceCube Neutrino Observatory, located at the geographic South Pole, consists of a cubic kilometer of ice instrumented with 5,160 \acp{DOM} \cite{DOMs} on 86 strings, placed at depths between  \SI{1450}{\meter} and \SI{2450}{\meter}.
The main detector array consists of 78 strings arranged in a roughly hexagonal array, with an average horizontal distance of 125 m between neighboring strings \cite{kiryluk2008neutrino} (see \cref{fig:deepcore_illustration}).  Each string supports 60 DOMs separated vertically by  \SI{17}{\meter}.
 Each \ac{DOM} contains a 10"~\ac{PMT} facing downwards. 
Around the center string in the deepest part of the array where the optical transparency of the ice is highest, modules on 8 additional strings have been installed. This volume, named  ``DeepCore''~\cite{DeepCore}, has an increased spatial density of DOMs and features \acp{PMT} with an enhanced \ac{QE} compared to the main array.
The IceCube Observatory is constructed to detect neutrino interactions spanning the energy range of a few GeV to several PeV, with the purpose of exploring properties of both the cosmos and fundamental properties of neutrinos \cite{Halzen:2010yj, DeepCore}.

Charged particles resulting from neutrino interactions in the ice will emit Cherenkov radiation detectable by IceCube. The number of detected photons is many orders of magnitude lower than those emitted. For the neutrinos detectable by IceCube, the signal in a neutrino event may range from a low of a few to a high of order 100,000 detected photoelectrons.
In the energy range this work is focusing on---from a few GeV up to around 100\,GeV---a typical interaction deposits about 49 GeV of energy and has just 17 detected \cite{retro} photoelectrons.
This signal, however, is interspersed with noise stemming from, for example, radioactive decays in the DOMs. Event candidates are read out from the detector at a trigger rate of around \SI{2.7}{\kilo\hertz}~\cite{icecube_detector}.
These recorded events are dominated by triggers due to downgoing atmospheric muons, followed by first random triggers caused by noise pulses, second by atmospheric neutrinos at a rate less than 10\,mHz ~\cite{DeepCore}. 

\subsection{The Challenge of IceCube Event Classification and Reconstruction}

Event reconstruction can be framed as a problem of parameter inference. Given a set of detector observations, the reconstruction aims to infer the physics properties of a (neutrino) interaction. A reconstruction algorithm defines and implements this process of taking input data and returning parameter estimators.

\paragraph{Parameters of Interest and Categorization of Events}
The parameters of interest estimated in the reconstruction vary by application.
The deposited energy, the direction of the neutrino candidate, and the interaction vertex are of central relevance to many IceCube physics analyses. Events are also categorized into multiple morphological classes, of which only two are relevant for this work because the vast majority of the sample is below 100 GeV.
These two event classes serve as proxies for the underlying neutrino flavor and interaction type. ``Track-like'' events are proxies for $\nu_{\mu}$ charged-current (CC) interactions from atmospheric and astrophysical neutrinos. These events contain a muon that can travel a long distance inside the detector while emitting Cherenkov radiation, producing a signature that looks like a track going through the ice. However, track-like signatures can also be produced by atmospheric muons from cosmic rays, and from the 17$\%$ of $\nu_{\tau}$ CC interactions that produce $\tau$ leptons that decay into muons\cite{Zyla:2020zbs}, but these are not considered true tracks for the purposes of this work.
The other class comprises ``cascade-like'' events, containing everything that is not described by the track-like class. These events consist of electromagnetic and hadronic particle showers, such as those produced in $\nu_{e, \tau, \mu}$ neutral-current (NC) and $\nu_{e,\tau}$ CC interactions.
For brevity, we will refer to the classification of track ($\mathcal{T}$) and cascade ($\mathcal{C}$) morphologies as ``\trackvscascade''.
Another classification task is the discrimination of neutrino events from atmospheric muons events, which will be referred to as ``\nuvsmu''.
\paragraph{Input Data}
Data from the IceCube PMTs is digitized with waveform digitizers which record the majority of the PMT signals from neutrino interactions. These waveforms undergo an unfolding process \cite{wavedeform} that attempts to extract estimated photon arrival times and the corresponding charge of individual photoelectrons each of which is a so-called "pulse". The pulses form the detector response to an interaction and make up a time series. Each element in this sequence corresponds to the time $t$ at which the \ac{PMT} readout indicates a measured pulse with charge $q$.
Most pulses in the energy regime considered in this work stem from single photons. These result in pulse charges close to one photoelectron, with variation in the charge due to the stochastic nature of the PMT amplification process. The detection of multiple, nearly instantaneous photons results in higher per-pulse charges.
The \acp{PMT} themselves are located at fixed \ac{DOM} positions ($D_{\text{xyz}}$) and have an empirically determined \ac{QE}. The DOMs deployed in DeepCore have an efficiency that is roughly a factor of 1.35 times the \ac{QE} of standard DOMs \cite{DeepCore}. These six variables of the time sequence are summarized in \cref{tab:features} and form the input data for the reconstruction.

\begin{table}[h]
\centering
\caption{Node features in graph representation of neutrino events. The units shown here are before the preprocessing mentioned in \cref{sec:preprocessing}.}
          \label{tab:features}
    \begin{tabular}{lll}
    \hline
    Feature & Description & Unit  \\ 
    \hline  
     $D_\text{xyz}$ & Position of \acp{DOM} in IceCube coordinates & m\\
     $t$ & Pulse time relative to trigger time & ns \\
     $q$ & Charge of a pulse & P.E. \\
     $QE$ & Quantum efficiency of \ac{PMT} & - \\
     \hline
    \end{tabular}

\end{table}

The amount of Cherenkov radiation produced in an event depends primarily on the energy of the interacting particle, and less energetic neutrinos often lead to a reduced amount of  Cherenkov light. Consequently, the number of pulses for an event, $n_\text{pulses}$, is highly dependent on the event itself, and makes low energy neutrinos particularly difficult to identify and reconstruct. 
In this study, we focus on neutrinos with energy less than \SI{1}{\tera\electronvolt}. The range of \SIrange{1}{30}{\giga\electronvolt} is of particular importance to the study of atmospheric neutrino oscillations \cite{atmospheric_neutrino_oscillation, tau_appearance}.
For typical events in this energy range, $n_\text{pulses}$ can range from below ten to several hundred after noise hit removal.
Because some pulses in an event are due to noise pulses, low-energy neutrino events often suffer from a relatively poor signal-to-noise  ratio. In addition, at low energies the 
track events can be so short that they cannot be easily distinguished from cascade-like events. While the irregular geometry of IceCube help distinguish events that would otherwise leave identical detector responses in a regular geometry configuration, it provides an additional layer of complexity to the development of reconstruction algorithms. In addition, reconstruction is further complicated by varying ice properties as a function of x,y,z, position in the IceCube array \cite{photoelectrons}. The z-dependence of the ice absorption and effective scattering is shown to the left in \cref{fig:deepcore_illustration}.

\subsubsection{Traditional Reconstruction Methods}

Maximum likelihood estimation is a standard technique for parameter inference. The key ingredients are the likelihood itself and a maximization strategy. The exact reconstruction likelihood for IceCube events is, however, intractable and can only be approximated.
In IceCube, event reconstructions range in complexity from simple analytical approximations of likelihoods \cite{SPE} to highly sophisticated detector response functions based on photon ray tracing \cite{retro}.

A major challenge in modeling the detector response is including the right amount of detail of an event's Cherenkov light emission profile, as well as the inhomogeneous optical properties of the ice, which is visualized on the left side of \cref{fig:deepcore_illustration}. Both affect the expected photon pattern in the \acp{PMT}. Analytic approximations, which are fast but inaccurate, are predominantly used as first guesses for both online and offline processing \cite{SPE}. More sophisticated reconstruction techniques come with a higher computational cost and can only be applied to a subset of the events. For the neutrino energies considered in this work, (i.e.\ low-energy events between 1 \,GeV -- 1000 \, GeV) such likelihood-based reconstructions were used, for example, in the analyses of Refs.~\cite{atmospheric_neutrino_oscillation, tau_appearance}.
We will use the state-of-the-art IceCube low-energy reconstruction algorithm \retro \cite{retro} as a benchmark for our work.
On average, \retro requires around 40 seconds to reconstruct one event on a single CPU core, which is one of the main shortcomings of the method. 
Another limitation is its use of simplifying approximations, such as the assumed uniformity in the azimuthal response of the IceCube \acp{DOM}, as well as the assumption that ice properties change as a function of depth only.

The IceCube Upgrade ~\cite{ishihara2019icecube} will augment IceCube's detection capabilities by adding additional detector strings featuring new DOM types: the ``mDOM'' and ``DEgg''. The mDOMs carry 24 3"~\acp{PMT} providing almost homogeneous angular coverage, while the DEggs carry two 8" \acp{PMT}, one facing up and one down \cite{mdom_degg}.
Adjusting \retro to work with the IceCube Upgrade is difficult while keeping memory usage and computing time at a reasonable level, due to assumed symmetries being broken and the increase of different module types.

\subsubsection{Machine-Learning-Based Reconstruction Methods}

An alternative method to maximum likelihood estimation is regression with \acp{NN} \cite{springer_neural_networks}. Instead of approximating the likelihood and traversing its parameter space with an optimization algorithm, a regression algorithm returns parameter estimates directly.
These regression models are trained by minimizing a defined loss function. Since an optimal regression algorithm may be highly nonlinear, artificial \acp{NN} can offer a viable solution.
A regression algorithm needs to map the ragged input data of shape $[n_\text{pulses}, 6]$, where the number of columns refers to the six node features ($D_x$, $D_y$, $D_z$, $t$, $q$, $QE$) described in \cref{tab:features},  onto an output of shape $[1,D]$ estimating event level truth information, where $D$ is the number of parameters we are interested in.
The spatio-temporal nature of IceCube data, with the addition of varying sequence lengths, makes it difficult to map event reconstruction in IceCube to common machine learning techniques.
Previously published IceCube machine learning methods embed events into pseudo-images and perform regression using  a \ac{CNN} \cite{MircoDNN}. This class of algorithms is effective at identifying local features in data — i.e., CNNs efficiently group similar data points in the compressed ``latent space’’ representation of the input images. However, embedding IceCube events into images is a lossy operation aggregating pulse information into per-\ac{DOM} summary statistics, a process that severely degrades the information available in low-energy events. Due to this reason, the \ac{CNN} reconstruction method is mainly used at energies higher than what is considered in our work, but an adaptation for our energy range exists \cite{Micallef_2021} \cite{Yu_2021}.
An alternative \ac{NN} architecture employing a \ac{GNN}~\cite{gnn_paper} approach is used in \cite{berkley_paper}, but focuses solely on $\nu/\mu$ classification.

We propose a general reconstruction method based on \acp{GNN} that can be applied to the entire energy range of IceCube, is compatible with the IceCube Upgrade, and can reconstruct events studied in this work at speeds fast enough to run in real time online processing at the South Pole. 
We consider a set of reconstruction attributes: $\nu/\mu$ classification, \tc classification, deposited energy $E$, zenith and azimuth angles ($\theta$, $\phi$), and $x$, $y$, and $z$ coordinates of the interaction vertex, denoted by $V_\text{xyz}$.
We will benchmark our reconstruction method on a simulated low-energy IceCube sample used for atmospheric neutrino oscillation studies. 

\section{Graph Neural Networks Applied to IceCube Data}

Generally, \acp{GNN} are \acp{NN} that work on graph representations of data. A graph consists of {\it nodes} interconnected by {\it edges}. Nodes are associated with data, and the edges specify the relationship among the nodes. By adopting graphs as the input data structure, the idea of convolution generalizes from the application of filters on the rigid structure of grids to abstract mathematical operators that utilize the interconnection of nodes in its computation. This allows such models to naturally incorporate an irregular geometry directly in the edge specification of the graph, without imposing artificial constraints. For this reason, GNNs are a natural choice of machine learning paradigm for problems with irregular geometry, such as IceCube event reconstruction.

We choose to represent each event by a single graph. Each observed pulse is represented by a node in the graph and contains the per-DOM information shown in \cref{tab:features}. Each node in the graph is connected to its 8 nearest neighbors based on the Euclidean distance, and for this reason we consider the interconnectivity of the nodes in the graphs to be spatial \cite{spatial_spectral}.

\subsection{Preprocessing of IceCube Events}
\label{sec:preprocessing}
The variables listed in \cref{tab:features} come in different units and can span over different ranges, for example
the relative positions of triggered DOMs can take values between tens of meters to $>1$~km; or
the times spanned by pulses may range from tens of nanoseconds to several microseconds. While neural networks can in theory process data in any range of real numbers, the complexity of the loss landscape that a model navigates during training is highly dependent on the relative scale of the input data. In addition, distributions not centered around zero can lead to a slower convergence time \cite{LeCun2012}. Each input variable $x$ is therefore transformed into $\bar{x}$ using
\begin{equation}
    \bar{x} = \frac{x - {P}_{\text{50th}}(x)}{{P}_{\text{75th}}(x) - {P}_{\text{25th}}(x)}, 
\end{equation}
where ${P}_{i\text{th}}(x)$ is the $i$-th percentile of the distribution of input feature $x$. This transformation brings the input variables into roughly similar orders of magnitude, gives a median of zero and makes them unitless.

\subsection{Model Architecture }
Our GNN implementation, henceforth referred to as \gnn, is a general purpose reconstruction and classification algorithm that is applied directly to the event pulses from the IceCube detector. Therefore, the algorithm does not rely on the pulses to be aggregated into summary statistics like \cite{MircoDNN}. Instead, \gnn uses graph learning to extract features from the pulses, making the mapping of measurements to prediction a fully learnable task. Additionally, \gnn learns the optimal edges between nodes and can be applied to events with any number of pulses. \gnn is implemented using GraphNeT~\cite{graphnet}, a software framework for graph neural networks in neutrino telescopes built using PyTorch Geometric~\cite{pytorch_geometric}.\\

\gnn uses a convolutional operator {\it EdgeConv}~\cite{edgeconv}, developed to act on point cloud graphs in computer vision segmentation analysis. For every node $n_j$ with node features $x_j$, the operator convolves  $x_j$ via local neighborhood of $n_j$ as
\begin{equation}
    \tilde{x}_j = \!\! \sum_{i=1}^{N_{\mbox{\tiny neighbors}}} \!\! \text{MLP}(x_j, x_j-x_i),
    \label{eqn:edge_conv}
\end{equation}
\noindent
where $\tilde{x}_j$ denotes the convolved node features of $n_j$, and the \ac{MLP} takes as input the unconvolved node features of $n_j$ and the pairwise difference between the unconvolved node features of $n_j$  and its $i$-th neighbor. Thus, the connectivity of the node in question serves as a specification of which neighboring nodes contribute to the convolution operation.
A full convolution pass of the input graph consists of repeating \cref{eqn:edge_conv} for every node in the graph. When compared to convolutions as understood from CNNs, \cref{eqn:edge_conv} corresponds to a convolutional operator where the kernel size is analogous to $N_{\mbox{\tiny neighbors}}$ and the "pixels" on which the kernel acts are defined by the edges, instead of the position of the kernel.\ 
 
\begin{figure}[H]
    \centering
    \begin{center}

    \includegraphics[width=\textwidth]{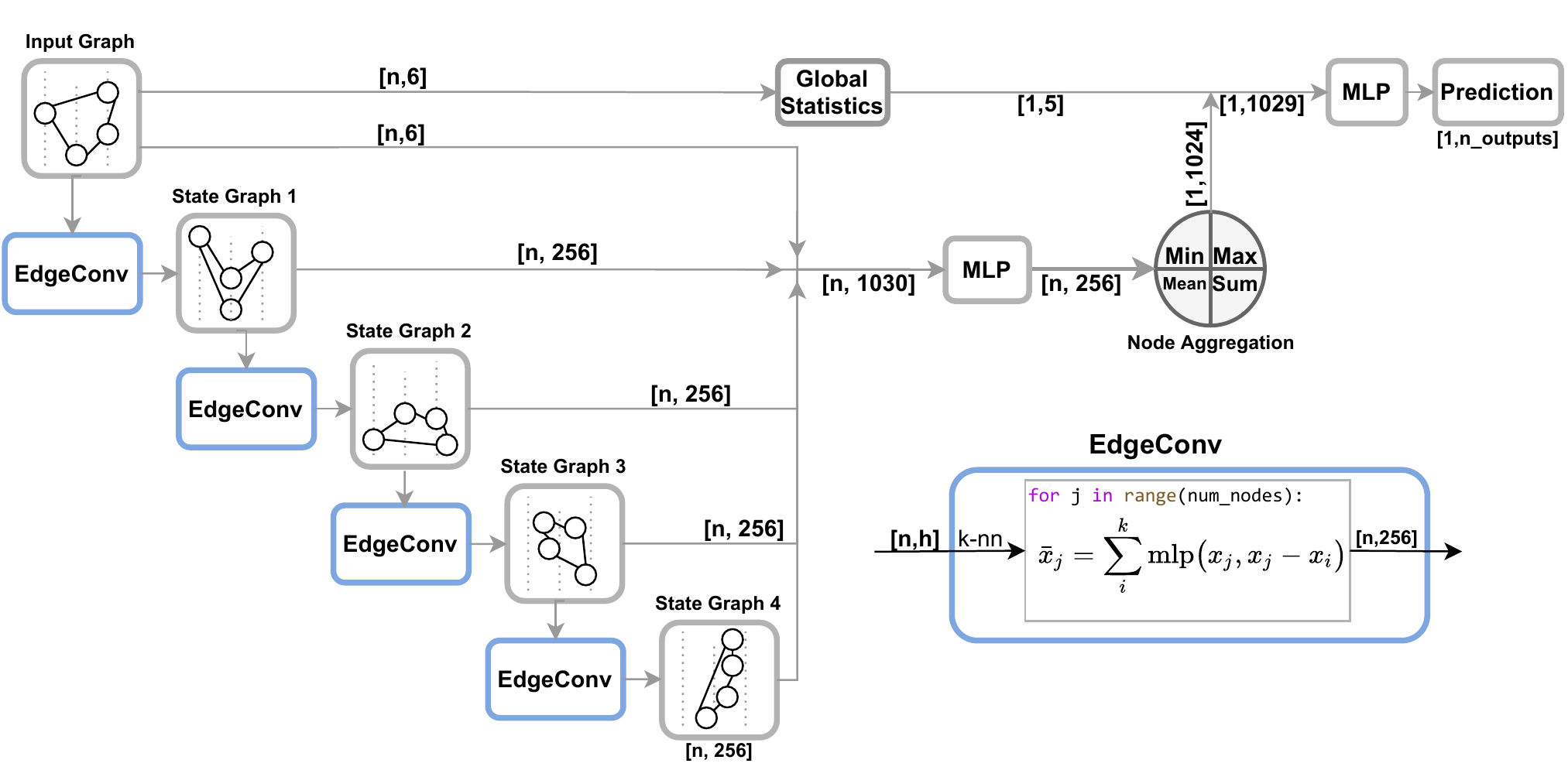}
    \caption{A diagram of the architecture of \gnn. The "Input Graph" is a toy illustration of the input, and the subsequent "State Graphs" illustrates how the node position and connectivity change after convolutions by EdgeConv.  An illustration of the inner workings of the EdgeConv block is provided to the right, where h represents an arbitrary number of columns. The number of pulses is represented by $n$.}
    \label{fig:AlgoOverview}
    \end{center}
\end{figure}

 The full architecture of \gnn is shown in \cref{fig:AlgoOverview}. First, the following 5 global statistics are calculated from the input graph: node homophily ratio \cite{homophily} of $D_\text{xyz}$ and $t$, and number of pulses in the graph. The homophily ratio is the ratio of connected node pairs that share the same node feature, and thus a number between 0 and 1. The homophily ratio of $D_\text{xyz}$ indicates what fraction of connected pulses originate from the same PMT. For a graph where all nodes are connected to each other,  a value of 0.5 indicates that half the pulses in the event are same-PMT pulses. As such, the homophily ratio quantifies how many of the pulses originate from the same PMT and how many of the pulses happened at the same time. Second, the input graph is propagated through 4 different EdgeConv blocks such that the output from one flows into the next. As illustrated in the bottom right corner of \cref{fig:AlgoOverview}, the EdgeConv block is initialized with a \ac{knn} computation that redefines the edges of the input graph. This step lets \gnn calculate the 8 nearest Euclidean neighbors and means that the first convolution block connects the nodes in true $xyz$-space and the subsequent convolution blocks connects the nodes in an increasingly abstract latent space. By letting the subsequent convolutional blocks interpret the convolved $xyz$-coordinates from the past block, \gnn is allowed to connect the nodes arbitrarily in each convolutional step, which effectively lets \gnn learn the optimal neighborhoods of the event for a specific task given the 8-nearest-neighbors constraint.

Third, the input graph and each state graph are concatenated together into a $[n,1030]$-dimensional array that is passed through a fully connected \ac{MLP} block containing two layers. The block maps the $[n,1030]$-dimensional array into a $[n, 256]$-dimensional array. This array is aggregated node-wise into summary statistics in four parallel ways; \textit{mean}, \textit{min}, \textit{max}, and \textit{sum}, each corresponding to a many-to-one projection of the form $f: [n, 256] \longrightarrow [1,256]$\footnote{While more sophisticated pooling methods were tested, none improved upon this choice.}. Aggregations are concatenated together to minimize information loss, which produces an array with dimension $[1,4 \cdot 256]$ that is concatenated together with the initially calculated 5 global statistics, producing a $[1, 1029]$-dimensional input array to the subsequent 2-layer \ac{MLP} block that makes the final prediction by mapping the $[1,1029]$-dimensional array to a $[1, {n}_\text{outputs}]$-dimensional output. This node aggregation scheme removes the need for zero-padding of the input data, and allows the model to function on any number of pulses.

The architecture shown in \cref{fig:AlgoOverview} is the result of multiple iterations of architecture tests. We find that increasing the number of convolutional layers leads to identical performance but at increased training time, whereas a decrease in the number of convolutional layers leads to a noticeably worse performance. In addition, a wide variety of data-driven pooling operations have been tested, but none improved upon the many-to-one projections shown in \cref{fig:AlgoOverview}. Also, hyperparameters such as the number of nearest neighbors used in the k-nearest-neighbors computation have been subject to tuning. We find that for $k$ larger than 8, the convolutions become too coarse to learn local features, whereas for $k$ smaller than 8 the convolutions become too fine to be globally descriptive. Also, the internal dimensions of the data referenced both in text and in \cref{fig:AlgoOverview} have been subject to hyperparameter optimization. These dimensions effectively set the number of learnable parameters in the model. We find that a significant reduction in learnable parameters yields an under-parameterized model that cannot learn the complicated relationships between data and task. A significant increase in learnable parameters increases training time and yields similar performance as reported for the current configuration.

\noindent

\subsection{Training Configuration and Loss Functions}

A \gnn network is trained for each of the  reconstruction variables: deposited energy $E$, zenith and azimuth angles ($\theta$, $\phi$), the interaction vertex $V_\text{xyz}$, $\nu/\mu$ classification and \tc classification. This totals 6 independently trained models. The difference between each model is the choice of loss function, number of outputs, and training selection.

Each model is trained with a batch size of 1024, using the \textsc{ADAM}~\cite{adam} optimizer. The batch size indicates the number of events in each sub sample of the training set, on which the gradients of the loss are calculated. A custom piece-wise-linear implementation of a One-Cycle \cite{one_cycle} learning rate schedule with a warm-up period \cite{warmup} is used. The scheduler increases the global learning rate linearly from $10^{-5}$ to $10^{-3}$ during the first 50\% of iterations in the first epoch, and thereafter linearly decreases the learning rate to $10^{-5}$ during the remaining iterations in the 30 epoch training budget. To counteract overfitting, early stopping \cite{earlystopping} is implemented with a patience of 5 epochs. 

For all classification tasks, the Binary Cross Entropy \cite{cross-entropy} loss is used, since we only consider two categories: neutrino or muon events, and tracks or cascades. However, for each regression task, a specific loss function is chosen.

For regression of deposited energy, a LogCosh \cite{logcosh} function is used
\begin{equation}
    \text{Loss}(E) = \ln \left( \cosh \big( R_E\big) \right),
\end{equation}

\noindent
applied to the residual $R_E = \log_{10}(E_{\text{reco}} / \text{GeV}) -  \log_{10}  (E_{\text{true}} / \text{GeV})$.
The embedding ${f}: E_{\text{true}} \longrightarrow {\log}_{10} (E_{\text{true}} / \text{GeV})$ is added to account for the large range that the deposited energy spans, and $\log_{10}(E_{\text{reco}} / \text{GeV})$ is the model's prediction of the embedded, deposited energy. LogCosh is symmetric around zero and, therefore, punishes over- and under-estimation equally. When compared to more conventional choices such as mean-squared error (MSE), LogCosh offers a steadier gradient around zero and is approximately linear for large residuals, both beneficial to the training process.  \\

\begin{table}[h]
\centering
\caption{Targets of the GNN-based classification and reconstruction algorithm, based on the truth values of the event simulation. Includes definitions of residual distributions for regression targets.}
          \label{tab:targets}
    \resizebox{\columnwidth}{!}{\begin{tabular}{lll}
    \hline
    Targets & Description & Residual Definition \\ 
    \hline  
     $\nu/\mu$ & Classification of neutrino vs. muon events     &  --    \\
     $E$             & Deposited energy of neutrino interaction       & ${R}_{{E}} =\text{log}_{10}({E}_\text{reco}) - \text{log}_{10}({E}_\text{true}) $ \\
     $\theta, \phi$  & Zenith and azimuth angles of neutrino         & ${R}_{\text{angle}} = \text{angle}_\text{reco} - \text{angle}_\text{true}$\\
     $\vec{r}$     & Direction vector of neutrino & ${R}_{\vec{r}} = {\arccos} \frac{\Vec{r}_\text{reco} \cdot \Vec{r}_\text{true}}{|\Vec{r}_\text{reco}||\Vec{r}_\text{true}|}$ \\
     $V_\text{xyz}$           & Vertex position of neutrino interaction  & ${R}_{V_\text{xyz}} = |\Vec{{P}}_{\text{reco}} - \Vec{{P}}_{\text{true}}|$ \\ 
     $\mathcal{T}/\mathcal{C}$ & Classification into tracks and cascades  & --  \\
     \hline
    \end{tabular}}
\end{table}

For angular regression, a von Mises--Fisher \cite{von_mises} Sine-Cosine loss is used, where the true angle $\text{\o}$ is embedded into a 2D vector space using ${f}: \text{\o} \longrightarrow \rvect{\sin\text{\o}, \cos\text{\o}}$. \gnn is then tasked with predicting this embedded vector together with an uncertainty \textit{k}. The d=2 von Mises-Fisher distribution, where d is the dimension of the unit vectors, is given by
$
    p(\bar{x}|\bar{u},k) = C_{2}(k)\exp{(k \bar{u}\cdot \bar{x})} = C_{2} \exp{(k\cos{\Delta\text{\o}})}
    \label{eqn:von_mises}
$
where $\bar{x}$ is the predicted embedded vector; $\bar{u}$ is the embedding of the true angle; $k$ resembles ${1}/{\sigma^2}$ of a normal distribution; and $C_{2}$ is a normalization constant written in terms of modified Bessel functions.
The d=2 von Mises--Fisher distribution describes a probability distribution on a 1-sphere embedded in $\mathbb{R}^{2}$ and bears resemblance to loss functions such as $1 - \cos(\Delta\text{\o})$ but with the added functionality of uncertainty estimation via $k$. Note that $\Delta \text{\o}$ is chosen to be the angle between the predicted and the true embedding vector \footnote{A more intuitive approach such as a d=3 von Mises-Fisher, where $\Delta \text{\o}$ is chosen to be the angular difference between predicted and true direction vectors in  $\mathbb{R}^3$, was found to lead to suboptimal results because the azimuthal component of the loss is too dominant at lower energies. Substantial improvement in zenith reconstruction is gained by estimating the two angles separately.}. The loss function is created by taking the negative log of $p(\bar{x}|\bar{u},k)$:
\begin{equation}
    \text{Loss}(\text{\o}) = -\ln{p(\bar{x}|\bar{u},k)} = -\ln\big(C_{2}(k)\big) -\ln{(k/4\pi)} + k + \ln{(1-e^{-2k})} - k\cos{(\Delta \text{\o})}.
    \label{eqn:von_mises_Loss}
\end{equation}
After zenith and azimuth angles are regressed individually, the direction is produced by transforming zenith and azimuth into a direction vector $\vec{r}_\text{reco} \in \mathbb{R}^3$.

\indent The true interaction vertex is embedded using ${f}: {(x,y,z)} \longrightarrow {\left(\frac{x}{|\max(x)|},\frac{y}{\max(y)|},\frac{z}{|\max(z)|}\right)}$ for the same reasons mentioned in \cref{sec:preprocessing}. \gnn then predicts the embedded interaction vertex vector, and the loss function is the Euclidean distance between the true ($\tilde{{{P}}}_{\text{true}}$) and reconstructed ($\tilde{{{P}}}_{\text{reco}}$) embedding vectors.

\section{Performance on Low-Energy Neutrino Events}
\label{sec:performance}

In this section, we quantify the performance of our proposed algorithm based on simulated data in the energy range of 1\,GeV--1000\,GeV, where the majority of selected events are below 100\,GeV (see \cref{fig:truth_distributions}),  and compare it with the state-of-the-art \retro algorithm. When possible, we provide comparisons for tracks ($\nu_\mu CC)$ and cascades (all other neutrino interactions) separately.  At the end of the section, a short examination of the runtime performance will be provided.

\subsection{Selected Datasets}
\label{subsec:data_selection}

The IceCube simulation used for training and testing the \gnn classifications and reconstructions is borrowed from the collaboration's simulation for neutrino oscillation analyses. The dataset and selection process is similar to the ones described in \cite{tau_appearance}, which is also used in \cite{retro}. The interactions were simulated with GENIE \cite{Andreopoulos_2010}, and simulations of the propagation of secondaries, particle showers, and the propagation of Cherenkov light are the same as used in \cite{tau_appearance}.
The event selection aims to provide a clean neutrino sample in the DeepCore region of IceCube by removing pure noise events, atmospheric muon events and by applying containment cuts. \retro is run at the late stages of the event selection (level 7 in \cite{tau_appearance}) because of it's high computational cost. We have chosen this late stage of the event selection as our dataset because it allows for a direct comparison with \retro.\\
The resulting selection totals approximately 8.3 million simulated neutrino and muon events, and because the event selection removes virtually all pure noise events, these have been omitted from this work. Once weighted to a physical spectrum, the data sample contains less than 5\% atmospheric muons \cite{tau_appearance} and the neutrino sample consists of mostly contained interactions below 100 GeV. Distributions of a few event level variables of the selected events are shown in \cref{fig:truth_distributions}. 
From the 8.3 million event sample, we construct three task specific datasets, summarized in \cref{tab:data}.

\begin{table}[h]
\centering
\caption{Selected datasets for the reconstruction and classification tasks for \gnn. Testing sets are defined as the data on which no back-propagation has been made, and therefore includes the validation set.}
\resizebox{\columnwidth}{!}{\begin{tabular}{l|l|lllllll}
    \hline
    {\bf Task}        & Total  &{ \bf $\nu_{e}^{NC}$}(\%)&{ \bf $\nu_{e}^{CC}$} (\%)& { \bf $\nu_{\mu}^{NC}$}(\%)& { \bf $\nu_{\mu}^{CC}$}(\%)& { \bf $\nu_{\tau}^{NC}$}(\%)& { \bf $\nu_{\tau}^{CC}$}(\%)& { \bf $\mu$}~(\%)  \\
    \hline
    $\nu/\mu$ (Train)                    & 215k & 1.6 & 15.0 & 1.6 & 15.0 & 4.4 & 12.3 & 50.1 \\
    $\nu/\mu$ (Test)                     & 106k & 1.6 & 15.1 & 1.5 & 15.3 & 4.5 & 12.2 & 49.8 \\
    Reconstruction (Train)               & 3.76M & 3.2 & 30.0 & 3.1 & 30.3 & 9.0 & 24.4 & \\
    Reconstruction (Test)                & 4.37M & 1.4 & 12.8 & 6.6 & 64.3 & 4.0 & 11.0 & \\
    $\mathcal{T}/\mathcal{C}$ (Train)    & 731k & 16.7 & 16.7 & 16.7 & 49.9 & \\
    $\mathcal{T}/\mathcal{C}$ (Test)     & 7.4M & 0.8 & 21.2 & 3.9 & 48.2 & 7.0 & 18.9 & \\

    \hline
\end{tabular}}
\label{tab:data}
\end{table}
Because the event selection aims to produce a pure neutrino sample, the raw count of muon events in the 8.2 million sample is less than 200k. In addition, there are significantly more cascade events than tracks ($\nu_\mu$ CC), making the full data set class-imbalanced. This work under-samples~\cite{undersampling} the full data set into three task-specific training sets with corresponding test sets to counteract the imbalance between the three classes: tracks, cascades and muons. 

The $\mathcal{T}/\mathcal{C}$ training data set contains an even amount of  $\nu_\mu$CC events (considered "tracks") and CC $\nu_e$ + NC events (considered "cascades"), and due to the high number of cascade events, the training sample consists of only 731k events. The $\nu_\tau$ events are omitted from the training sample (but included in the test sample), since 17\% of $\nu_\tau^{CC}$ events produce track-like signatures that may confuse the model during training with a track-like signature but a cascade-like label. The corresponding test set for  $\mathcal{T}/\mathcal{C}$ constitutes the remaining neutrino events, mostly consisting of cascades.  For the $\nu/\mu$ classification task, the training dataset is chosen such that there are an equal amount of neutrino and muons events. Because of the few muon events, this dataset is the smallest, and the neutrinos have been sampled such that there is an even amount of each flavor.

For reconstruction, a dataset with equal amounts of neutrino flavors has been chosen.  On this dataset, energy, zenith, azimuth, and interaction vertex are regressed. In order to keep the balance between tracks and cascades, one may most naturally choose the  dataset from the $\mathcal{T}/\mathcal{C}$ task, because it is balanced between the two event classes, but this selection yields significantly lower statistics. It was found that the high statistics choice provided general improvements compared to reconstructions from \gnn trained purely on the balanced $\mathcal{T}/\mathcal{C}$ set.

The classification results are available in \cref{fig:neutrino_classification} and reconstruction results are shown in \cref{fig:neutrino_reconstruction}. The normalized distributions of regression targets on test and training sets from \cref{tab:targets} are plotted in \cref{fig:truth_distributions} for each event selection in \cref{tab:data}.  The $\mathcal{T}/\mathcal{C}$ and Reconstruction selections are similar in target distributions, whereas the $\nu/\mu$ event selection have additional artifacts due to the presence of muons.

\subsection{Event Classification Performance}

\begin{figure}[h]
        \begin{center}
        \includegraphics[scale=0.7]{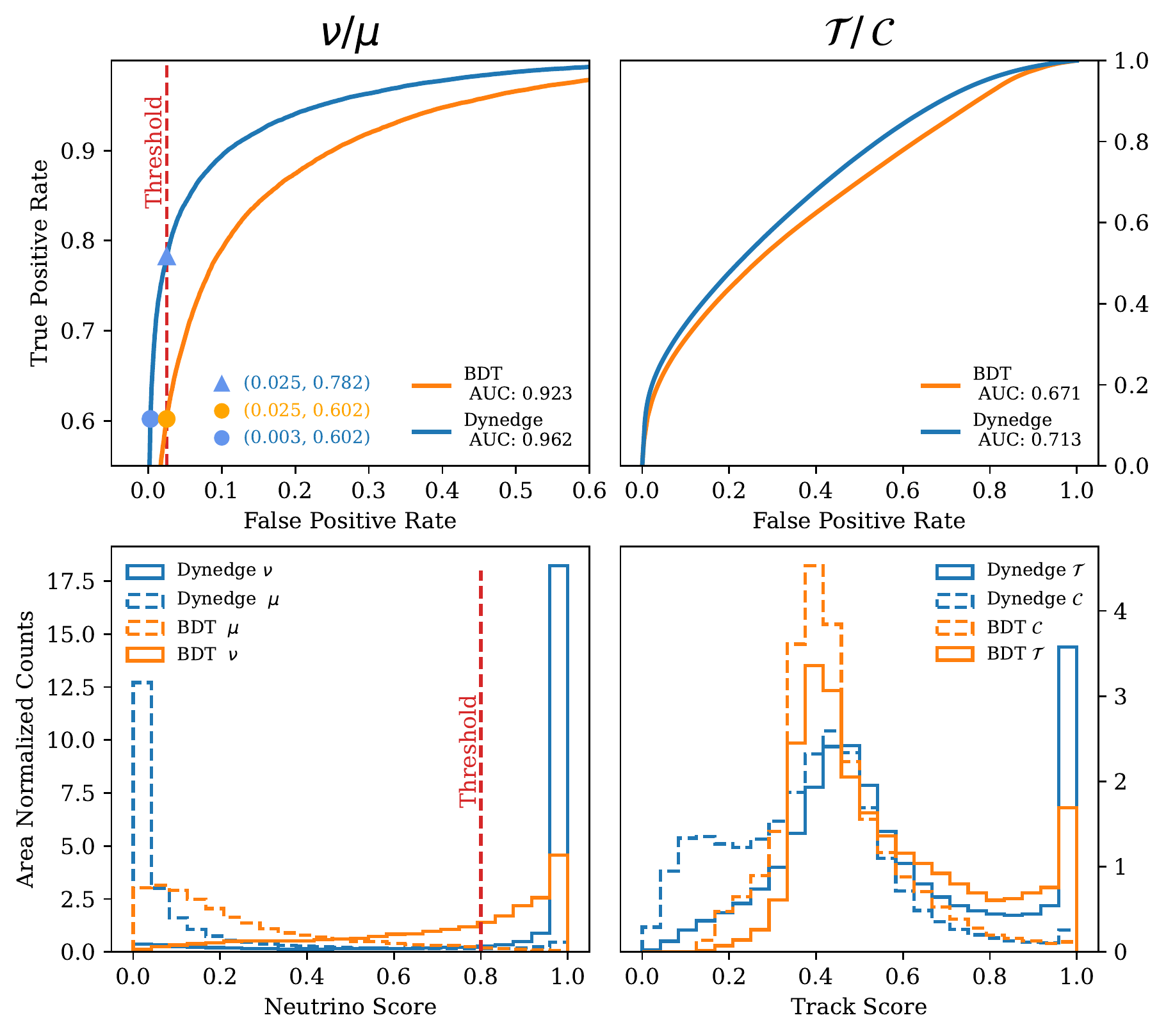}
        \caption{{\bf Left:} ROC curves for \gnn and current BDT $\nu/\mu$ classification models and below are the distribution of classification scores shown with the threshold labeled in red. {\bf Right:} ROC curve for \gnn and current BDT  $\mathcal{T}/\mathcal{C}$ classification models. Distributions in prediction scores are shown below.}
        \label{fig:neutrino_classification}
        \end{center}
\end{figure}

The performance of the \gnn classifiers and the currently-used \ac{BDT} methods is characterized using \ac{ROC}~\cite{ROC} curves, from which the \ac{AUC}~\cite{auc} is calculated.  The $\nu/\mu$ classification task is used in event selection, and for this reason a threshold for the classification scores, shown in the bottom left plot of \cref{fig:neutrino_classification}, is used to make a decision on whether an event is labeled a neutrino or a muon. A typical threshold in classification score is indicated in red in the left panel of \cref{fig:neutrino_classification}. Intersection points between the red line and the ROC curves represents the corresponding False Positive Rate (FPR) and True Positive Rate  (TPR)  at this choice of selection.

\noindent
By comparing the TPR of intersection points, one can deduce that \gnn offers increased signal efficiency by 18\% relative to the BDT at a FPR of 0.025. Alternatively, the FPR can be reduced by over a factor of 8 from 2.5\% to 0.3\% relative to the BDT at a fixed TPR of 0.6 for the $\nu/\mu$ classification task.
From the ROC curve for the \tc classification task, \gnn offers an increased AUC score of about 6\%. A threshold for the \tc classification task is not shown because the \tc classification scores, shown in the bottom right plot, are typically binned when used in an analysis. When inspecting the \tc classification score distributions in the bottom right of \cref{fig:neutrino_classification} it can be seen that \gnn offers a better separation of \tc events than the BDT. 

The large difference in classification performance shown in \cref{fig:neutrino_classification} between the \ac{BDT} and \gnn can be explained by the fact that classical \acp{BDT} require sequential input data, such as IceCube events, to be aggregated from $[{n}_\text{pulses}, {n}_\text{features}]$ into $[1,d]$-dimensional arrays as a preprocessing step, where $d$ is the number of event variables for the \acp{BDT} to act on. This preprocessing step reduces the information available to the \acp{BDT}.

\subsection{Event Reconstruction Performance of \gnn}
\label{subsec:eventreco}

\begin{figure}[H]
    \begin{center}
    \includegraphics[width = \textwidth]{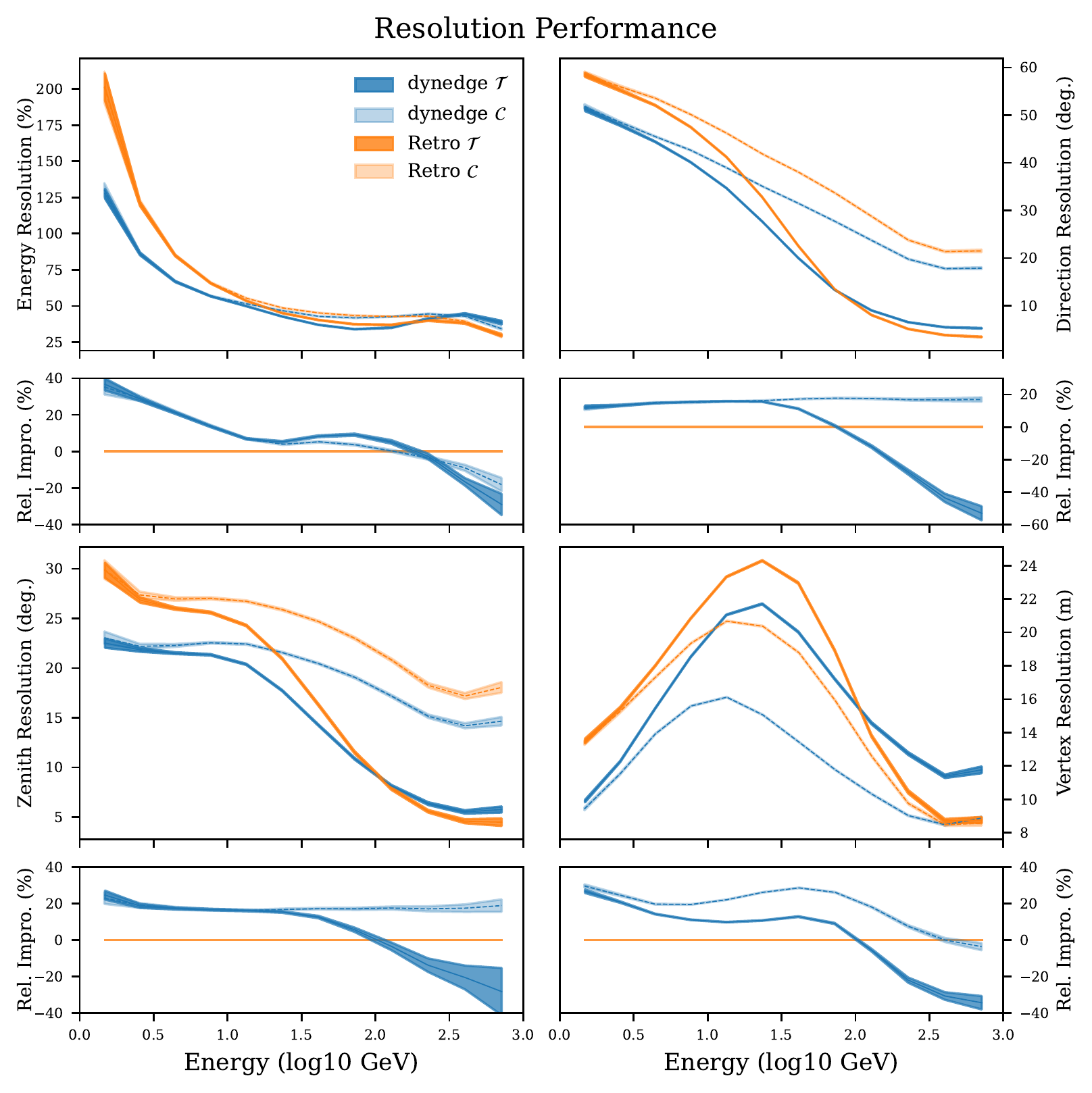}
    \caption{Event reconstruction performance for \gnn , estimating deposited energy (top left), zenith angle (top right), direction (bottom left), and interaction vertex (bottom right). In all four cases, the performance is compared to \retro, and the ratio below the plots shows relative improvement of \gnn w.r.t.\ \retro, where the light blue and dark blue curves represents the relative improvement for cascades and tracks, respectively. Positive values indicate an improvement in resolution. Lines represent reconstruction resolution, and the bands cover $1\sigma$ resolution uncertainty.} 
    \label{fig:neutrino_reconstruction}
    \end{center}
\end{figure}

The event reconstruction performance for deposited energy, zenith, direction, and interaction vertex is shown in \cref{fig:neutrino_reconstruction}. The performance metric is the resolution, which we quantify here as the width $W$ of the residual distribution $R$. For the deposited energy and zenith reconstructions, W is defined as
\begin{equation}
     {W} = \frac{{p}_{84\text{th}}\big({R}\big) - {p}_{16\text{th}}\big({R}\big)}{2}, 
    \label{eq:resolution}
\end{equation}
where ${p}_{16\text{th}}\big({x}\big)$ and ${p}_{84\text{th}}\big({x}\big)$ correspond to the 16th and 84th percentile. 
For a normal distribution, the quantile $W$ corresponds to the standard deviation $\sigma$, but it is more robust to outliers. For direction and interaction vertex reconstructions, (which have strictly positive residual distributions), the resolution $W$ is defined as the 50th percentile ${p}_{50\text{th}}\big({x}\big)$. Since reconstruction resolutions are highly dependent on the energy of the neutrino, we provide the resolutions binned in energy. Additionally, we separate the residual distributions in true track ($\nu_\mu$ CC) and cascade events to examine the performance of the algorithms in detail.

For comparison between \gnn and \retro, the relative improvement defined by
\begin{equation}
     {\text{relative improvement}} = \big( 1 - \frac{W_\text{\gnn}}{W_\text{\retro}}\big) \cdot 100
    \label{eq:rel_imp}
\end{equation}
is used. Here $W_\text{\gnn}$ and $W_\text{\retro}$ corresponds to the resolution of \gnn  and \retro for a given reconstruction task. The relative improvement provides a percentage comparison of the resolutions.

\begin{figure}[H]
        \begin{center}
        \includegraphics[width=\textwidth]{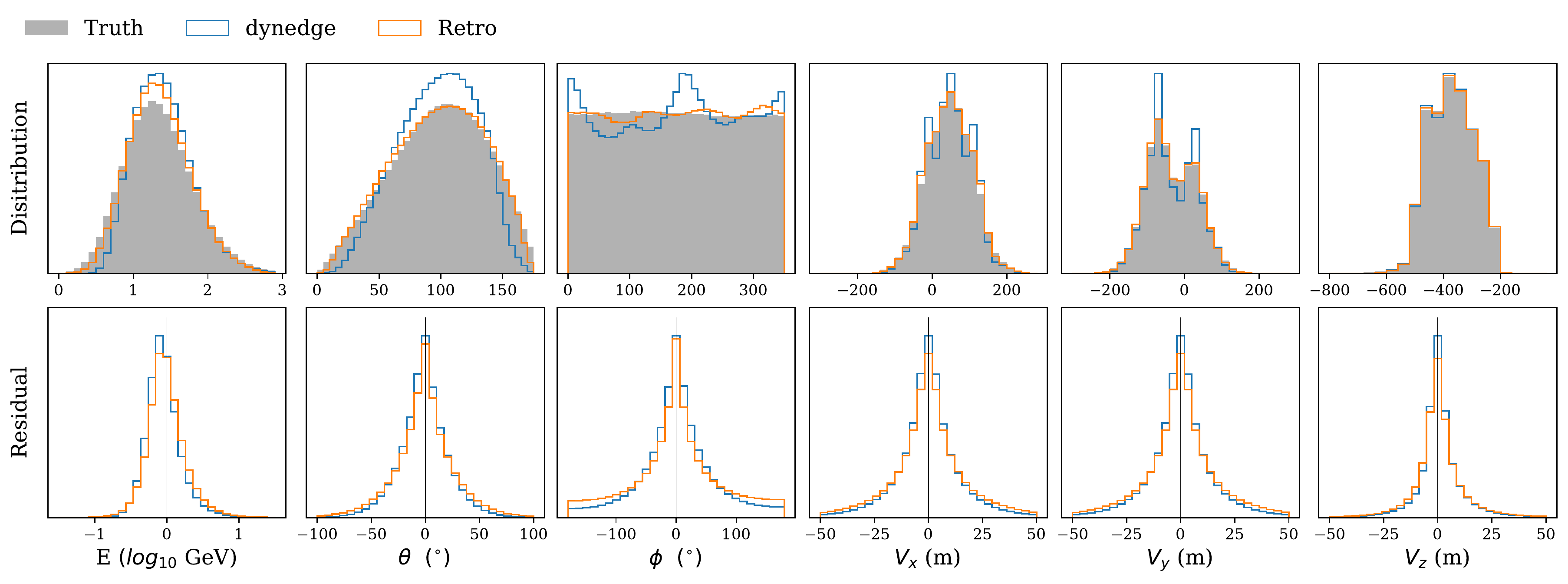}
        \caption{Area normalized distributions of predicted and true target variables on the reconstruction test set and their corresponding residual distributions. Residuals shown here follows the definitions shown in \cref{tab:targets} except for vertex coordinates, which show the relative difference between reconstructed and true coordinate. }
        \label{fig:truth_distributions}
        \end{center}
\end{figure}

The residual distributions for deposited energy (${R}_{{E}}$)\footnote{Notice that the residual distribution for energy is changed to ${R}_{{E}} = \frac{E_{\text{reco}} - E_{\text{true}}}{E_{\text{true}}}\cdot 100$ for the calculation of resolutions.}, angles (${R}_{\text{angle}}$), interaction vertex (${R}_{\text{xyz}}$), and direction (${R}_{\text{dir}}$) are defined in \cref{tab:targets} and some are shown in \cref{fig:truth_distributions}. From \cref{fig:truth_distributions} it can be seen that \gnn tends to over-estimate the deposited energy for very low-energetic events. This is partly due to the lower statistics and the relatively poor signal-to-noise ratio  below 30\,GeV.  On average, it will be optimal for a machine learning model to estimate a value close to the mean of the true distribution for examples where it has not yet learned a more optimal solution. Such examples can be considered to be difficult for the method, and by predicting a value close to the mean of the true distribution, the model minimizes the loss. This behavior is evident for \gnn in the reconstructions of zenith and azimuth, where events with a relatively high estimated uncertainty have a preferred solution that minimizes the loss function. For the regression of azimuth, this behavior yields multimodal artifacts due to the cyclic nature of the variable. As seen in \cref{fig:deepcore_illustration}, DeepCore have more strings in the north/south than east/west. The events with high estimated azimuthal uncertainty piled around $\pi$ and $2\pi$ are generally low energetic track events with a detector response similar to a neutrino traveling east-west, but are in fact traveling north or south. If there isn't enough information to pinpoint whether the event belongs in either north or south, the loss can be optimized by picking a value close to the observed locations of the pileups, as they correspond to a "mean" between north and south. The timing information of the event can be used to pick the optimal pile. We emphasize that while the distribution of predictions give the impression that \retro is closer to the true distribution than \gnn, the correct assessment of the quality of reconstructions comes from interpretations of the residual distributions. From the bottom panel of \cref{fig:truth_distributions} it's evident that \gnn produces reconstructions with  narrower residual distributions and therefore superior resolutions.  \\ \\
As can be seen from \cref{fig:neutrino_reconstruction}, \gnn performance  improves upon the existing \retro reconstruction in all 6 reconstruction variables for both track and cascade events between 1\,GeV -- 30\,GeV. This energy range is of particular importance to neutrino oscillation analyses, where flavor oscillations of atmospheric neutrinos appear below 25\,GeV \cite{tau_appearance}. The typical reconstruction improvement in this range is at the level of 15\%--20\%.
For zenith and direction reconstruction, the improvement is constant with energy for cascade-like events, while for track-like events the improvements ends around \SI{100}{\giga\electronvolt}.

\gnn performance relative to \retro generally decreases at higher energies, and is possibly due to the lower number of available training samples at these energies, as shown in \cref{fig:truth_distributions}. As mentioned in \cref{subsec:data_selection}, the reconstruction event selection is chosen to maximize available statistics with an equal amount of neutrino flavors, rather than optimizing the balance of track and cascade events. This choice improves on the overall performance under the given circumstances, but at the same time leads to an underrepresentation of track-like events, which further lowers the available statistics for this event type at higher energies.

\noindent

\subsection{Runtime Performance}
The inference speed test measures the wall-time required to reconstruct the zenith angle on 1 million neutrino events in batches of 7168 events\footnote{The inference speed was tested on a server running Ubuntu 20.04.3 LTS with a 64-core @2.00 GHz AMD EPYC 7662 using a single NVIDIA A100 SXM4 40 GB VRAM, 1TB of RAM, and 5TB of NVME disk space.}. The test includes the time required to load the batch into memory, apply re-scaling, convert the batch into graphs, move data to GPU memory, and pass them through \gnn. A budget of 40 parallel workers is allocated to feed the model with batches. The test does not include overhead associated with writing the predictions to disk. The test was repeated 5 times on both GPU and CPU. The average speed was found to be $30.6 \pm 2.0$\,kHz on GPU, and $0.221 \pm 0.004$\,kHz for a single CPU core. The inference speed was tested for other variables considered in this work and was found to be within the uncertainties of the above values. 
Since a separate \gnn model is trained for each variable, the individual reconstructions can be run in parallel, in which case the reconstruction rate of an event is equal to the inference speed if fully parallelized. If the reconstruction of each variable is not run in parallel,  the reconstruction rate of an event is approximately the inference speed divided by the number of desired variables. For a full reconstruction of energy, zenith, azimuth, interaction vertex, \nuvsmu, and \trackvscascade classification, the reconstruction rate is approximately 5.1\,kHz on the GPU, or 37\,Hz for the single CPU core, respectively. With classification and reconstruction speed of nearly double of the median IceCube trigger rate (2.7\,kHz), \gnn is in principle capable of real time reconstruction of IceCube events using a single GPU.

\section{Robustness Test}

The results presented in \cref{sec:performance} show that \gnn is suitable for both classification and reconstruction tasks in the low-energy range of IceCube, specifically for the neutrino-oscillation-relevant energy range. However, these results are computed on simulation based on a nominal configuration of the detector.  In  \cref{subsec:pertub}, we investigate the robustness of \gnn to perturbations of input variables $D_{xyz}$, $t$ and $q$. These variables constitute the node features, and as such the perturbation test probes the robustness of \gnn to systematic variations in the node features themselves. In  \cref{subsec:ice_variance}, we investigate the robustness of \gnn to changes in systematic uncertainties associated with the detector medium and the angular acceptance of the DOMs. Variations in such assumptions lead not to variations in the node features, but to the topology of the events. The test therefore probes the robustness of \gnn to variations in the connectivity of the graph representations of the events.

\subsection{Perturbation of Input Variables}
\label{subsec:pertub}
There are systematic uncertainties associated with the input variables shown in \cref{tab:features}. The position of each string is only known to a precision of a few meters horizontally, and the vertical position is known with a precision better than one meter~\cite{icecube_detector}\footnote{Strings are also not perfectly vertical, which results in a non-uniform uncertainty on the horizontal and vertical position, but this effect is neglected in this test.}. In this section, we investigate the variation in resolution and AUC from perturbations of input variables by $x \longrightarrow x + \epsilon$, where x is an input variable and $\epsilon$ is a random number from a normal distribution with standard~deviation~$\sigma_x$. 

\begin{figure}[H]
    \begin{center}
    \includegraphics[width = 1\textwidth]{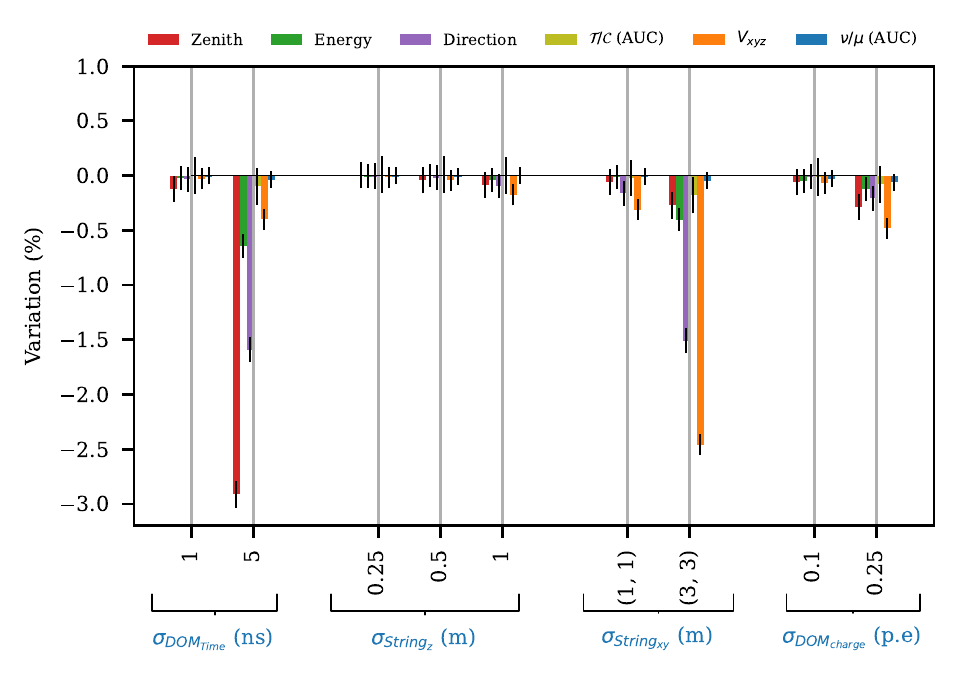}
    \caption{The variation in resolution and AUC induced by perturbation of input variables for \gnn. The standard deviations of the normal distributions from which the perturbations are drawn are shown in the x-axis, and the 4 different perturbation tests are labeled in blue at the bottom. Thin vertical lines centered on each bar are error bars.} 
    \label{fig:input_perturbation}
    \end{center}
\end{figure}
\noindent
For the \ac{DOM} positions, one $\epsilon$ is drawn for each string, resulting in correlated shifts for all DOMs in one string, and uncorrelated shifts between strings. The horizontal and vertical positions of the DOMs are also independently perturbed. Time and charge are perturbed pulse-wise, meaning that each pulse is perturbed independently. 
\gnn is not re-trained, i.e.\ we use the networks trained on nominal detector assumptions, and run on perturbed input datasets. We perturb time, $z$-coordinate, $xy$-coordinates, and charge separately, for a few choices of $\sigma$. We calculate the percentage variation with respect to the nominal resolution and AUC score. Our test addresses two questions. First, it gives an indication of the expected change in reconstruction performance due to a systematic shift in the input variables. Second, it serves as a gauge on feature importance for the different reconstruction tasks. 
In \cref{fig:input_perturbation}, variation in resolution and AUC is shown as a function of the perturbation width $\sigma$. A negative value indicates a \textit{worsening} with respect to the nominal performance. For example, -5\% means a worsened resolution with a width $W$ that is 5\% larger. For AUC scores, a variation by -5\% means a decrease of 5\%. As seen in \cref{fig:input_perturbation}, zenith is the most sensitive to perturbations of the time, as expected, since time perturbations of the pulses may reverse the direction. Perturbations of vertical and horizontal positions of strings have little effect on energy and direction, but impact the interaction vertex resolution. 

\subsection{Variations in Ice Properties and Module Acceptance}
\label{subsec:ice_variance}
The South Pole ice---IceCube's detector medium---is a glacier with an intricate structure of layers with varying optical properties, and the refrozen boreholes where strings were deployed are understood to have different optical properties than the bulk ice.
These sources of systematic uncertainty are constrained from fits to calibration data, but only to a finite precision of about $\pm 10\%$ of the nominal value \cite{atmospheric_neutrino_oscillation}. In the tests presented in this section, we quantify the robustness of our reconstruction to changes in the ice properties allowed by calibration data, using a five parameter model covering the most important sources of systematic detector uncertainties in IceCube.
\begin{figure}[H]
    \centering
    \includegraphics[width=0.6\textwidth]{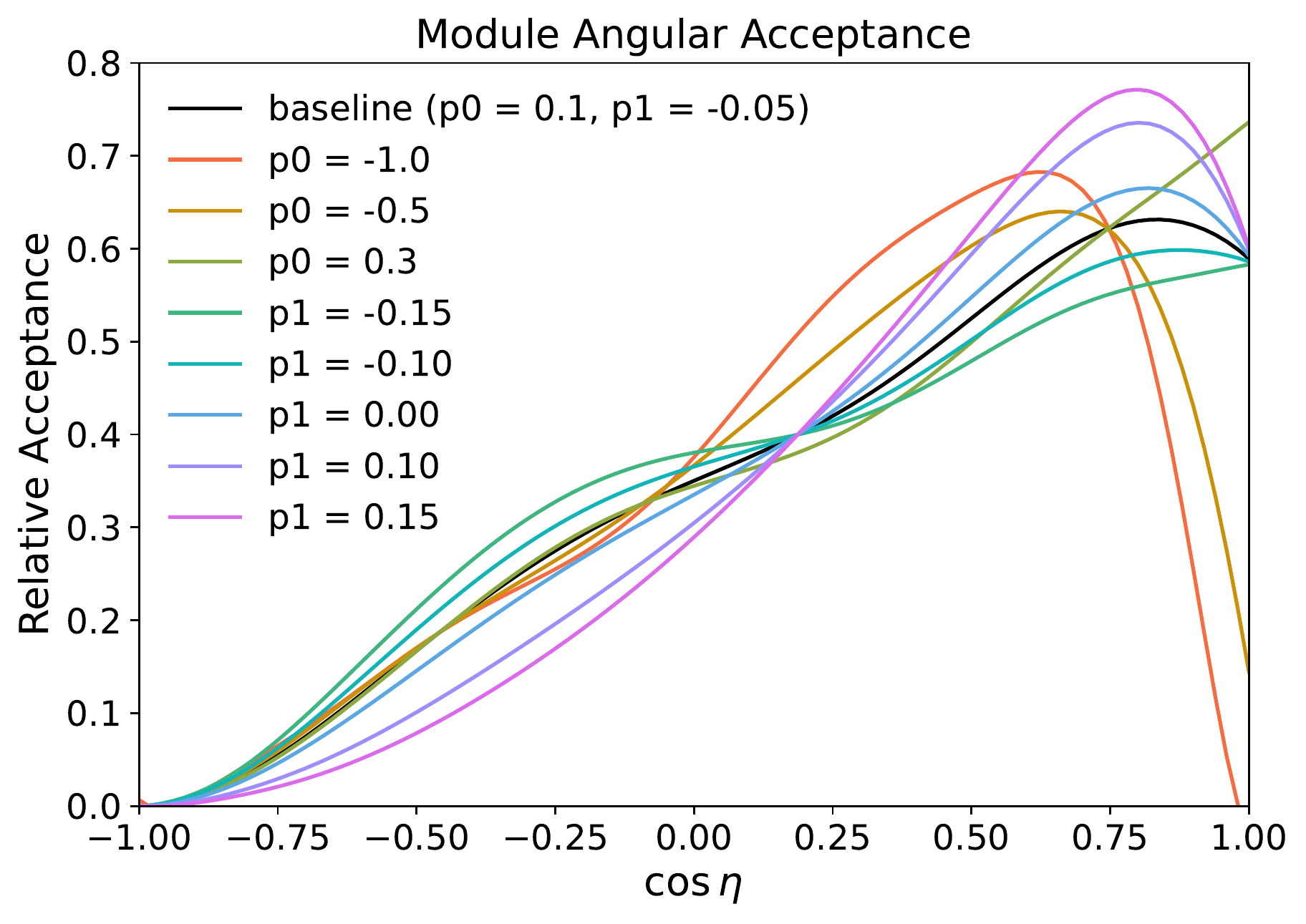}
    \caption{Variations in the \ac{DOM} angular acceptance used for the robustness tests. The angle $\eta$ is the photon arrival angle at the Module, with $\cos{\eta}=1$ being head-on towards the face of the \ac{PMT} and $\cos{\eta}=-1$ towards the backside of the \ac{PMT} where generally the acceptance drops to zero.}
    \label{fig:aa_curves}
\end{figure}
We consider 20 additional simulation sets  with altered ice properties and module acceptance. Specifically, 
there are 8 perturbed simulation sets varying the scattering  and absorption coefficients of the South Pole bulk ice by (-30\%, -10\%, +10\%, +30\%) independently, four sets with variations in the overall optical efficiency of \acp{DOM} (-10\%, -5\%, +5\%, +10\%), and 9 sets with independent variations in two key parameters $p_0$ and $p_1$ altering the angular acceptance of the DOMs. These last two parameters are the  principal components of a unification of different approaches in IceCube that model changes in the borehole ice via the angular photon acceptance\footnote{\url{https://github.com/icecube/angular_acceptance}}. As seen in \cref{fig:aa_curves}, an increase in $p_0$ 
leads to an increase in directly up-going photon acceptance (face of the \ac{PMT}), whereas changes in $p_1$ predominantly affect the photon acceptance around the waist of the \ac{PMT}.

In this subsection, we seek to quantify the robustness of \gnn to these variations  by letting the model predict on the systematic sets and comparing the bias, resolution, and AUC to that recorded on the nominal data set used in \cref{sec:performance}. For reference, the variation in the predictions of \gnn is compared with the variations in predictions from the current method \retro. For each individual systematic data set, only the events that are present in both the nominal set and the given systematic set are considered. The constraint to overlapping events is applied because the event selection process is itself sensitive to systematic uncertainties and can therefore lead to different distributions in parameters such as deposited energy, impacting event selection.

\subsubsection{Classification}
\begin{figure}[H]
    \begin{center}
    \includegraphics[width=\textwidth]{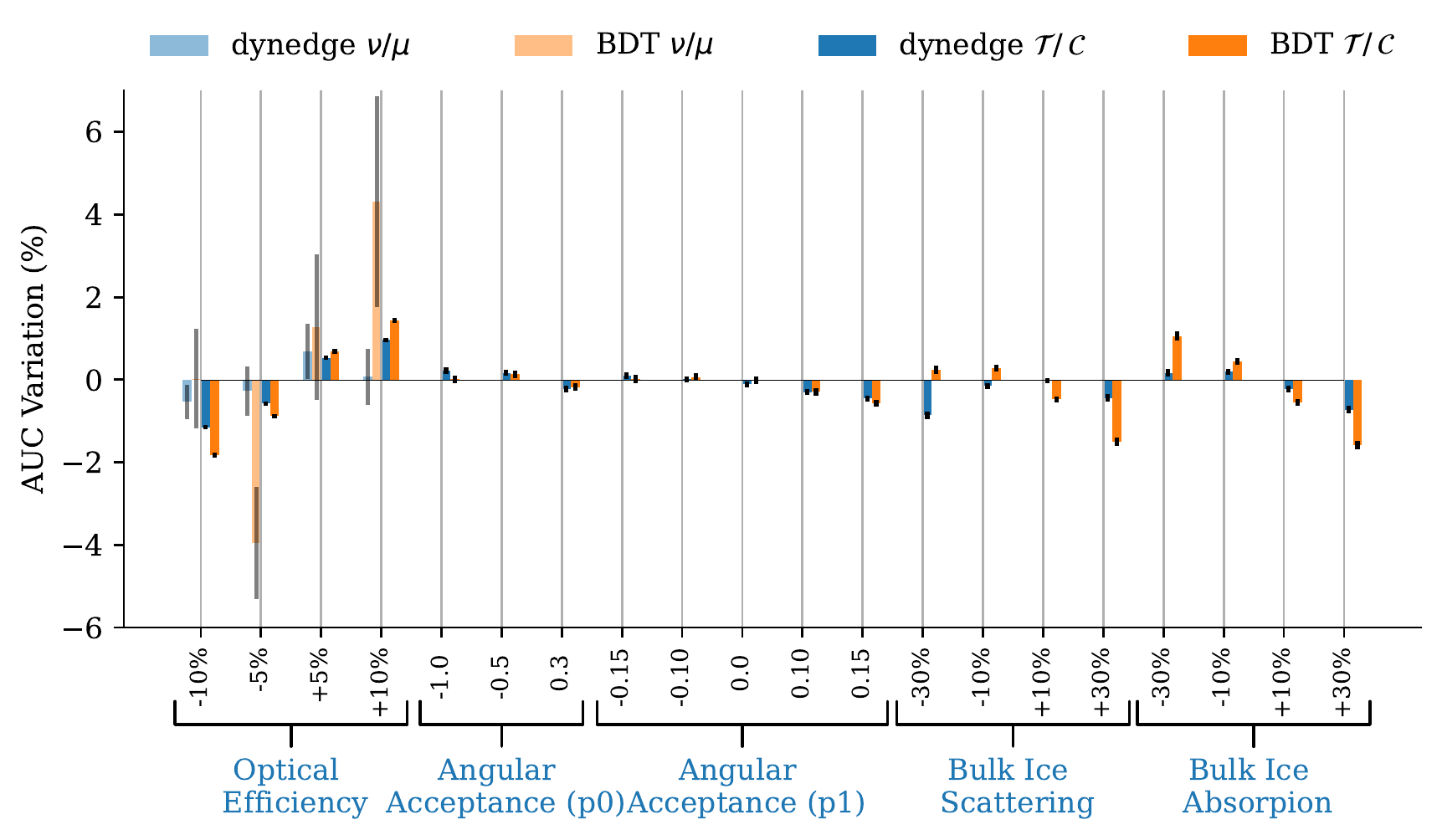}
    \caption[]{The variation in AUC for \gnn and the current BDT on the 20 different systematic sets. Track/cascade classification is denoted by \trackvscascade and neutrino/muon classification is denoted by \nuvsmu. Variation in \nuvsmu AUC is only shown on the first 4 samples, as the rest of the sets have no available muons.} 
    \label{fig:robustness_classification}
    \end{center}
\end{figure}

To quantify the robustness of the \trackvscascade and \nuvsmu tasks, the \ac{AUC} is computed from each ROC curve on each systematic set for both \gnn and the \acp{BDT}. The robustness of classification is then defined as the relative improvement in AUC as compared to the nominal AUC:
\begin{equation*}
    \text{AUC variation} = \left( 1 - \frac{\text{AUC}_\text{sys}}{\text{AUC}_\text{nom}}\right).
\end{equation*}
Currently, muon simulation is only available for the sets with variation in optical efficiency, limiting the test of \nuvsmu AUC variation to these sets. The test of \trackvscascade variation extends to all sets, and the results are shown in \cref{fig:robustness_classification}. 
 As seen in Fig.~\ref{fig:robustness_classification}, the error on AUC variation for $\nu/\mu$ is much higher than for $\mathcal{T}/\mathcal{C}$. This is due to the lower statistics of the data selection shown in Table~\ref{tab:data}. In general, we observe that $\mathcal{T}/\mathcal{C}$ classification has a variation well below $\pm 2$\% of the nominally expected AUC for both methods, \gnn and \retro.
  The  $\nu/\mu$ classification task, however, indicates a higher robustness of our proposed \gnn model than the current BDT classifier.

\subsubsection{Reconstruction}
The robustness of the reconstruction is reported in two main metrics; bias variation ($BV$) and resolution variation ($RV$), defined as ${BV}_\text{sys} = {p}_{50\text{th}}\big({R}_\text{sys}\big) -  {p}_{50\text{th}}\big({R}_\text{nom}\big)$  and ${RV}_\text{sys} = \left ( 1 - {{W}_\text{sys}}/{{W}_\text{nom}} \right)$ where $W$ and $R$ are as defined in \cref{subsec:eventreco}. These measures quantify the change relative to the nominal quantities. Since bias for the direction reconstruction is ambiguous, the bias variations of the zenith and azimuth reconstructions are shown instead.

\begin{figure}[H]
    \begin{center}
    \includegraphics[width=\textwidth]{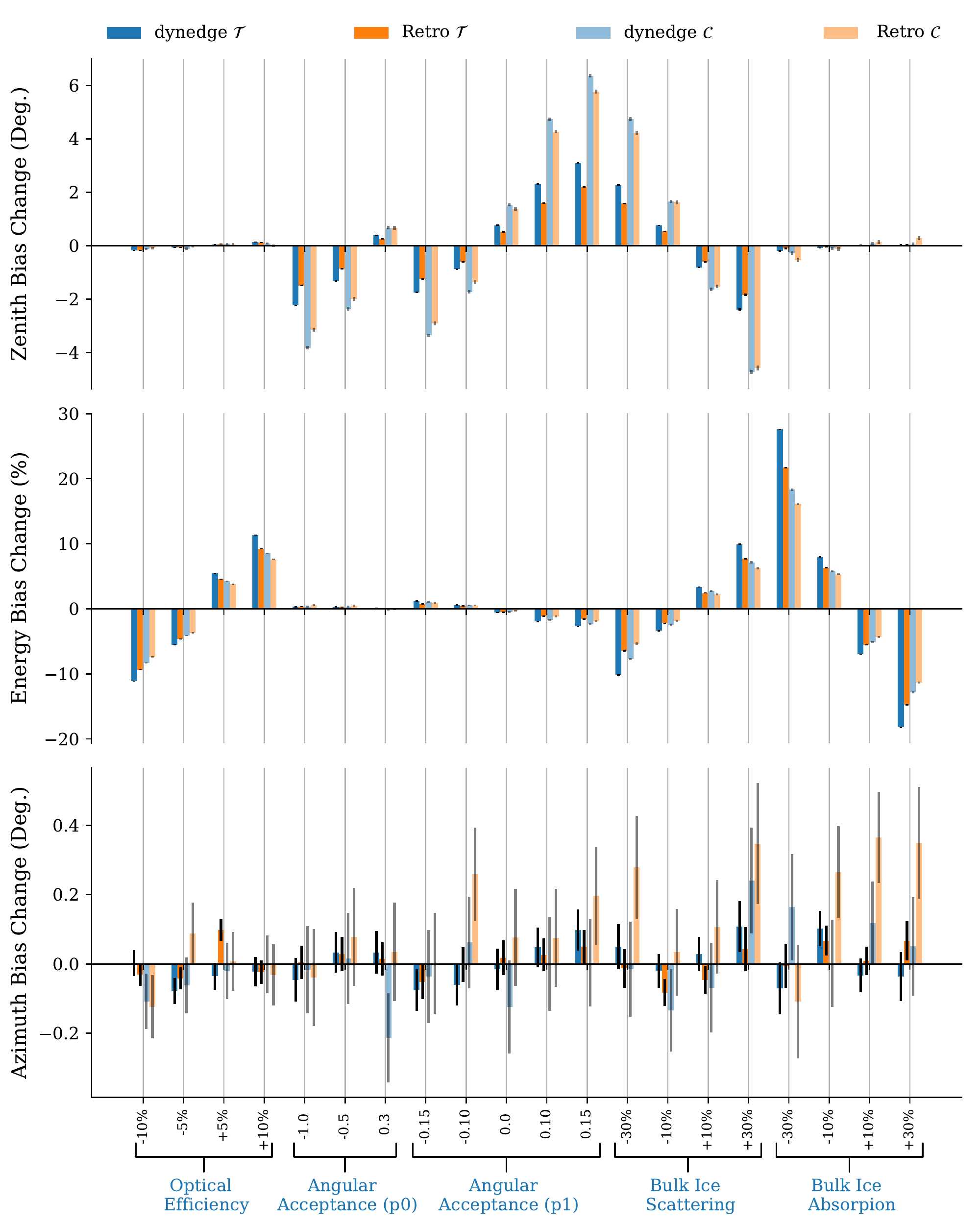}
    \caption[]{The variation in bias for \gnn and \retro on the 20 different systematic sets for reconstruction targets energy, zenith, and direction. Please notice that the y-axis values for azimuth is much smaller than for zenith.} 
    \label{fig:robustness_bias}
    \end{center}
\end{figure}
\noindent
\begin{figure}[H]
    \begin{center}
    \includegraphics[width=\textwidth]{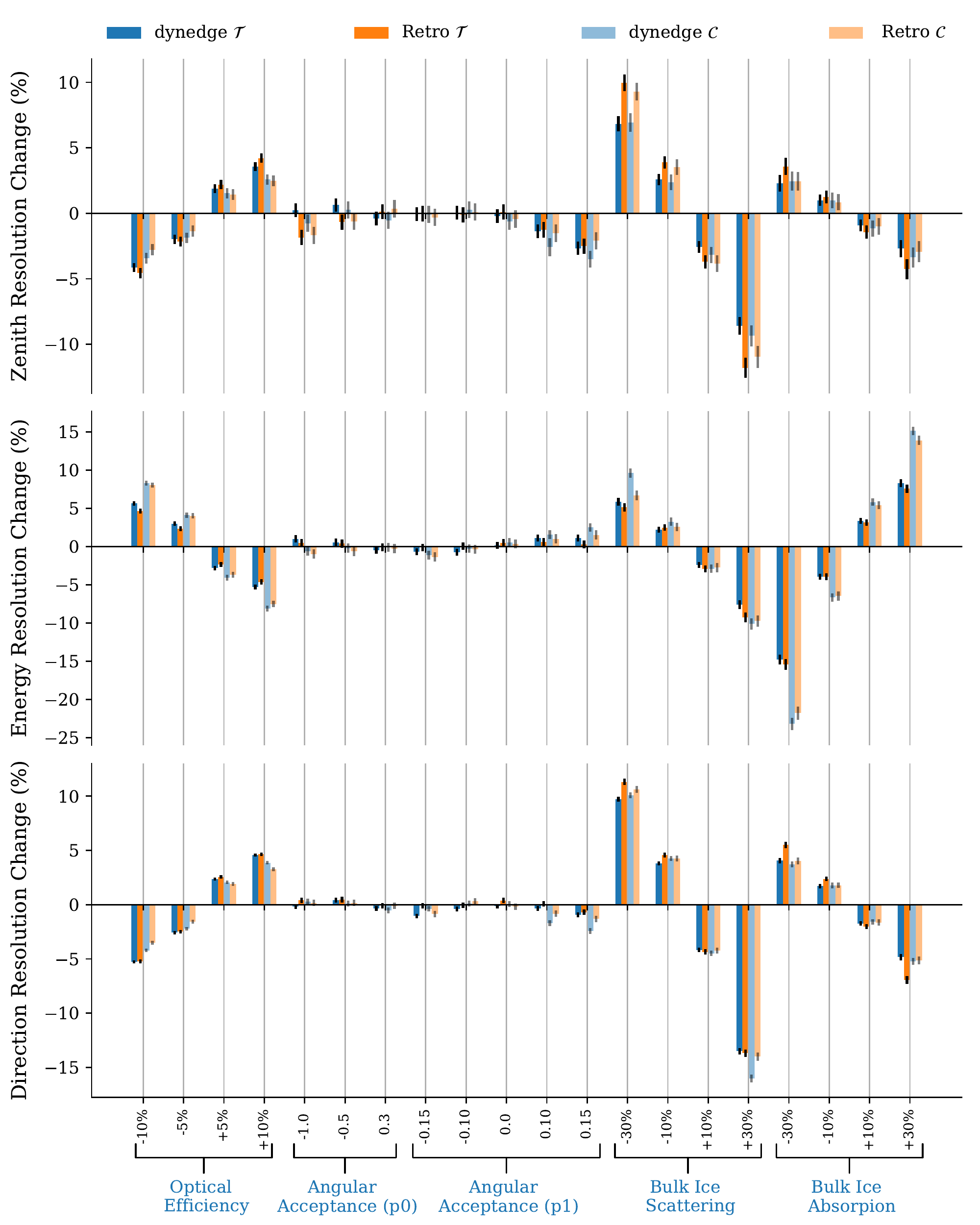}
    \caption[]{The variation in resolution for \gnn and \retro on the 20 different systematic sets for reconstruction targets energy, zenith, and direction.} 
    \label{fig:robustness_resolution}
    \end{center}
\end{figure}
\cref{fig:robustness_bias} shows the bias change for energy, zenith, and azimuth for the different simulation variations, while \cref{fig:robustness_resolution} reports the change in resolution.
The biases induced by changes in systematic uncertainties is nearly identical between \gnn and \retro for all variables considered, while a slight overall increase in magnitude for \gnn can be observed. Zenith bias is very sensitive to ice properties, whereas the induced bias in reconstructed energy is strongly affected by the optical efficiency and scattering/absorption of light, but less sensitive to angular acceptance of the DOMs. Such an effect is expected as the deposited energy is correlated with the number of measured pulses. Also, a linear change in DOM efficiency leads to an approximately linear change in energy bias for both methods. As seen in \cref{fig:robustness_resolution}, the overall effect on the resolution of the reconstruction is nearly identical for \gnn and \retro. While the zenith bias is unaffected by optical efficiency variations, the zenith resolution changes for both \gnn and \retro as a function of the optical efficiency. A decrease/increase in optical efficiency leads to a decrease/increase in available information for the methods, which, as expected, leads to a widening/narrowing of the residual distribution.

To summarize the difference in robustness between  \gnn and \retro or the current BDT classifiers, the RMS of the variations shown in \cref{fig:robustness_bias,fig:robustness_resolution,fig:robustness_classification} is presented in \cref{tab:robustness_rms}.       

\begin{table}[H]
\centering
\caption[]{Variation in bias, resolution, and AUC of reconstruction across all 20 systematic samples for \gnn and \retro for track and cascade events. (Results for $\nu/\mu$ classification are based only on the sets with varying \ac{DOM} optical efficiency.)}
\begin{tabular}{lcccc}

\hline
Target                             & \multicolumn{2}{c}{\gnn (RMS)} & \multicolumn{2}{c}{\retro/BDT (RMS)} \\
                                   & Tracks       & Cascades       & Tracks          & Cascades          \\ \hline
Energy Bias (\%)                   & 9.1          & 6.4            & 7.2             & 5.6               \\
Zenith Bias ($^\circ$)              & 1.4          & 3.2            & 1.0             & 2.4               \\
Azimuth Bias ($^\circ$)             & 0.07         & 0.09           & 0.04            & 0.19              \\ \hline
Energy Resolution (\%)             & 4.8          & 7.7            & 4.9             & 7.1               \\
Zenith Resolution (\%)             & 3.1          & 3.2            & 4.2             & 3.6               \\
Direction Resolution (\%)          & 4.5          & 4.9            & 4.9             & 4.5               \\
Vertex Resolution (\%)             & 1.2          & 1.2            & 3.2             & 5.3               \\ \hline
$\nu/\mu$ AUC (\%)                & \multicolumn{2}{c}{0.5}       & \multicolumn{2}{c}{3}               \\
$\mathcal{T}/\mathcal{C}$ AUC (\%) & \multicolumn{2}{c}{0.5}       & \multicolumn{2}{c}{0.8}             \\ \hline
\end{tabular}

\label{tab:robustness_rms}
\end{table}
The RMS values for \nuvsmu AUC are based only on the variations seen on 4 sets varying the optical efficiency of the \acp{DOM}. When considering the RMS values for the regression tasks, \gnn is generally affected by higher RMS values for both track and cascade events, indicating a higher variation induced by the systematic uncertainties compared to \retro. The single exception is the resolution in the reconstructed vertex, where \gnn exhibits a clear advantage over RETRO.

\section{Summary and Conclusions}

We propose a GNN-based reconstruction algorithm for IceCube events named \gnn, that can be applied to any IceCube event. We have selected simulated low-energy data used for studies of atmospheric neutrino oscillations in IceCube as our dataset, and in this energy range, we have benchmarked \gnn against the state-of-the-art reconstruction and classification algorithms as a proof of concept.

\gnn offers substantial improvements to both \trackvscascade and \nuvsmu classifications in the entire low energy range. In the energy range 1\,GeV--30\,GeV, relevant to standard atmospheric $\nu_{\mu}$ -- $\nu_{\tau}$ oscillation studies, \gnn exhibits a 15--20\,\% improvement in reconstruction of energy, zenith, direction, and interaction vertex. 
Outside the energy range relevant to neutrino oscillations, the improvement in event reconstruction decreases, and at high energies becomes worse than  \retro. This worsening effect is ascribed to lower statistics in the training set at these energies and a disproportional amount of cascades compared to more energetic (i.e. longer) tracks. Future studies will focus on improving performance in this region.

\gnn and \retro are both robust against systematic uncertainties in DOM optical efficiency and angular acceptance, as well as the scattering and absorption properties of the bulk ice. \gnn is also robust against random perturbations to inputs such as DOM position, timing, and charge.

Based on tests of reconstruction speed, \gnn could reconstruct events online at the South Pole. \gnn is also flexible enough to be compatible with the planned IceCube Upgrade featuring new DOM types on new strings, as well as other neutrino detectors with arbitrary geometries. This feature may make \gnn particularly useful for the next generation of large neutrino detectors, such as KM3Net \cite{km3net} and the proposed IceCube Gen2 \cite{gen2}. \gnn is implemented using GraphNeT~\cite{graphnet}.

\acknowledgments
The IceCube collaboration acknowledges the significant contributions to this manuscript from Rasmus Ørsøe. The authors gratefully acknowledge the support from the following agencies and institutions:
USA {\textendash} U.S. National Science Foundation-Office of Polar Programs,
U.S. National Science Foundation-Physics Division,
U.S. National Science Foundation-EPSCoR,
Wisconsin Alumni Research Foundation,
Center for High Throughput Computing (CHTC) at the University of Wisconsin{\textendash}Madison,
Open Science Grid (OSG),
Extreme Science and Engineering Discovery Environment (XSEDE),
Frontera computing project at the Texas Advanced Computing Center,
U.S. Department of Energy-National Energy Research Scientific Computing Center,
Particle astrophysics research computing center at the University of Maryland,
Institute for Cyber-Enabled Research at Michigan State University,
and Astroparticle physics computational facility at Marquette University;
Belgium {\textendash} Funds for Scientific Research (FRS-FNRS and FWO),
FWO Odysseus and Big Science programmes,
and Belgian Federal Science Policy Office (Belspo);
Germany {\textendash} Bundesministerium f{\"u}r Bildung und Forschung (BMBF),
Deutsche Forschungsgemeinschaft (DFG),
Helmholtz Alliance for Astroparticle Physics (HAP),
Initiative and Networking Fund of the Helmholtz Association,
Deutsches Elektronen Synchrotron (DESY),
and High Performance Computing cluster of the RWTH Aachen;
Sweden {\textendash} Swedish Research Council,
Swedish Polar Research Secretariat,
Swedish National Infrastructure for Computing (SNIC),
and Knut and Alice Wallenberg Foundation;
Australia {\textendash} Australian Research Council;
Canada {\textendash} Natural Sciences and Engineering Research Council of Canada,
Calcul Qu{\'e}bec, Compute Ontario, Canada Foundation for Innovation, WestGrid, and Compute Canada;
Denmark {\textendash} Villum Fonden, Carlsberg Foundation, and European Commission;
New Zealand {\textendash} Marsden Fund;
Japan {\textendash} Japan Society for Promotion of Science (JSPS)
and Institute for Global Prominent Research (IGPR) of Chiba University;
Korea {\textendash} National Research Foundation of Korea (NRF);
Switzerland {\textendash} Swiss National Science Foundation (SNSF);
United Kingdom {\textendash} Department of Physics, University of Oxford.

\bibliography{bibliography}
\bibliographystyle{JHEP.bst}

\end{document}